\documentclass[sigconf]{acmart}
\acmConference[SE 2030]{International Workshop on Software Engineering in 2030}{November 2024}{Puerto Galinàs (Brazil)}

\usepackage{multirow}
\usepackage{hyperref}
\usepackage{url}
\usepackage{subfig,graphicx}
\usepackage{multicol}
\usepackage{booktabs}
\usepackage{pifont}
\usepackage{bbding}
\usepackage{enumitem}
\usepackage{wrapfig}
\usepackage{svg}
\usepackage{listings}
\usepackage{xcolor}
\usepackage{makecell}
\usepackage{cleveref}
\usepackage{enumitem}
\usepackage{CJK}

\usepackage{tcolorbox}
\def\BibTeX{{\rm B\kern-.05em{\sc i\kern-.025em b}\kern-.08em
    T\kern-.1667em\lower.7ex\hbox{E}\kern-.125emX}}

\usepackage{xspace}

\makeatletter
\DeclareRobustCommand\bmvaOneDots{\futurelet\@let@token\bmv@onedotsaux}
\def\bmv@onedotsaux{\ifx\@let@token.,\else.,\null\fi\xspace}
\DeclareRobustCommand\bmvaOneDot{\futurelet\@let@token\bmv@onedotaux}
\def\bmv@onedotaux{\ifx\@let@token.\else.\null\fi\xspace}
\makeatother
\def\eg{\emph{e.g}\bmvaOneDots} 
\def\ie{\emph{i.e}\bmvaOneDots}

\def\etal{\emph{et al}\bmvaOneDot}

\begin{document}

\title{LLM for Mobile: An Initial Roadmap}

\author{Daihang Chen}
\affiliation{
  \institution{Beihang University}
  \country{China}}

\author{Yonghui Liu, Mingyi Zhou}
\affiliation{
  \institution{Monash University}
  \country{Australia}}

\author{Yanjie Zhao, Haoyu Wang}
\affiliation{
  \institution{Huazhong University of Science and Technology}
  \country{China}}

\author{Shuai Wang}
\affiliation{
  \institution{Hong Kong University of Science and Technology}
  \country{Hong Kong}}

\author{Xiao Chen}
\affiliation{
  \institution{The University of Newcastle}
  \country{Australia}}

\author{Tegawendé F. Bissyandé, Jacques Klein}
\affiliation{
  \institution{University of Luxembourg}
  \country{Luxembourg}}  

\author{Li Li}
\authornote{Corresponding author (lilicoding@ieee.org).}
\affiliation{
  \institution{Beihang University}
  \country{China}}

\begin{abstract}

% LLMs capabilities have been hugely assessed on various software engineering tasks. However, the capacity of LLMs to make mobiles intelligent has been less explored. 
% This is to our surprise since more intelligent mobile services could have a direct positive impact on the billions of mobile app users who interact daily with their smartphones.
% We therefore provide a research roadmap for
% guiding our fellow researchers in contributing to making the wedding of Mobile with LLMs a success. 

When mobile meets LLMs, mobile app users deserve to have more intelligent usage
experiences. For this to happen, we argue that there is a strong need to apply
LLMs for the mobile ecosystem. We therefore provide a research roadmap for
guiding our fellow researchers to achieve that as a whole. In this roadmap, we
sum up six directions that we believe are urgently required for research to
enable native intelligence in mobile devices. In each direction, we further
summarize the current research progress and the gaps that still need to be
filled by our fellow researchers.

\end{abstract}


\maketitle

\section{Introduction}

Large Language Model (LLM) has been the emergent buzzword in the SE community since the successful release of ChatGPT, a conversation-based AI system powered by OpenAI’s GPT-3.5 model, and the successful release of Copilot, GitHub’s AI developer tool supported by OpenAI’s Codex model.
It quickly becomes the most popular research topic in software engineering (if not in computer science).
The research efforts mainly focus on exploring two directions.
The first direction is related to applying SE methods to improve LLMs.
Indeed, as a new technique, LLM also comes with limitations that need to be resolved in order to apply LLMs in practice, as what has happened with other emerging technologies.
The other direction is to apply LLMs to resolve traditional SE tasks (e.g., code generation, unit test generation, etc.).
Our fellow researchers have experimentally shown that LLMs can achieve better results, compared to approaches that do not use AI or only adopt pre-LLM AI techniques.

Mobile Software Engineering (MSE) has been a hot research area in Software Engineering (SE).
It generally involves applying traditional software engineering methodologies (concepts, methods, tools, models, programming styles) to mobile software systems (such as Android or iOS) and apps, which are often distributed through app stores~\cite{li2017static, kong2018automated, li2019rebooting, zhan2021research, liu2022deep}.
So far, this hot research topic has attracted lots of attention from software engineering researchers who have subsequently made significant contributions to the MSE community from various aspects, such as Security and Privacy Analysis~\cite{li2015iccta, li2016droidra, hu2019dating, gao2019negative, li2017understanding, sun2022demystifying, samhi2022difuzer}, App Quality Assurance~\cite{sun2022mining, zhao2022towards, li2020cda, li2018cid, cai2019large}, App Store Analysis~\cite{obie2021first, wang2018beyond, wang2018android}, etc.

With the great results achieved by applying LLMs for SE and the flourishing mobile ecosystem, we believe it is time to apply LLMs for mobile. 
The smart devices will only be ``smart'' if LLMs are embedded.
At the moment, our fellow researchers have also seen the opportunities to apply LLMs for mobile and hence conducted several studies in this field.
However, the research roadmap for applying LLM for mobile has not yet been sketched.
To fill this research gap, in this position paper, we commit to summarizing the initial roadmap of applying LLM for mobile.

\begin{figure*}[!h]
    \centering
    \includegraphics[width=\linewidth]{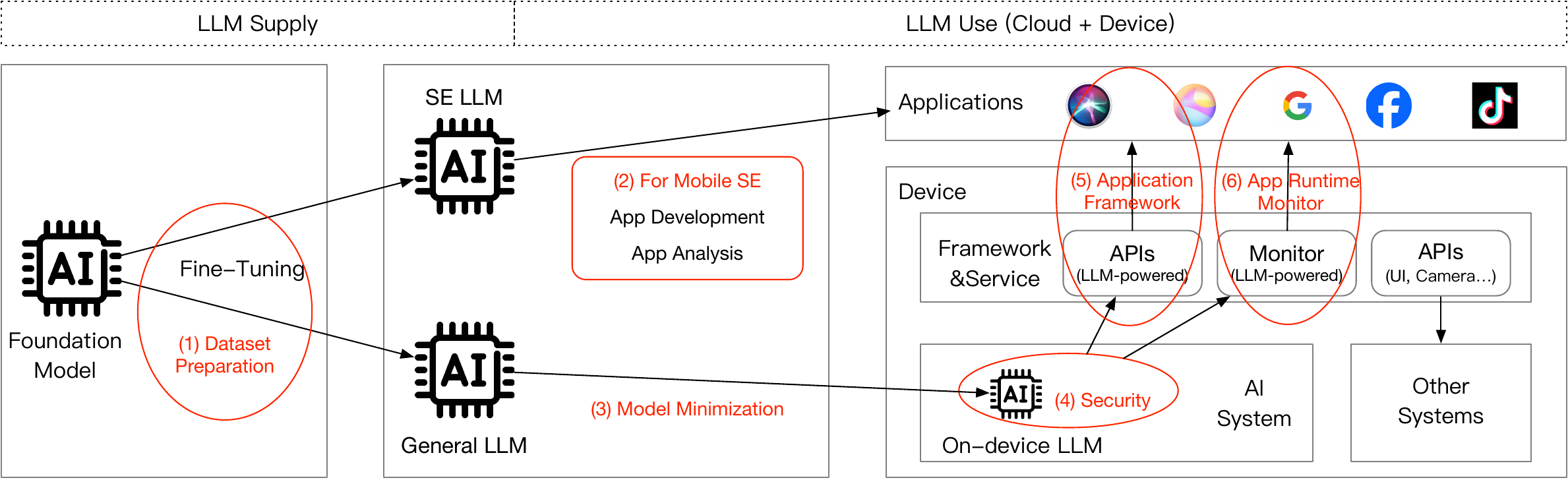}
	\caption{Roadmap in Applying LLMs for Mobile.}
    \label{fig:overview}
\end{figure*}

Figure~\ref{fig:overview} provides an overview of the roadmap.
In general, we divide the LLM for mobile tasks into two phases: LLM Supply and LLM Use.
The former phase involves preparing the right LLMs for solving downstream tasks, while the latter phase concerns the usages (or inference) of LLMs in mobile devices through local models (i.e., deployed in the device as part of the operating system) or online models (i.e., deployed in the cloud).
In these two phases, we further summarize six research directions that need to be further researched in order to seamlessly integrate LLMs into the mobile ecosystem. 
The six directions are depicted below.

\begin{itemize}
\item \textbf{Preparing datasets for fine-tuning LLMs dedicated to mobile.} In
particular, we advocate that the datasets should include User Experience (UX)
scenarios that enable better ways for apps to interact with LLMs, SE scenarios
that allow more efficient app development and analyses, and other multi-modal
data processing scenarios (e.g., sensor data, app logs) that enable LLMs to
process various types of data and tasks on mobile devices.
    
\item \textbf{Applying LLMs for mobile app development and analysis.} For app development, the whole lifecycle (i.e., requirement, design, coding, testing and debugging, maintenance, etc.) should be considered. For app analysis, both static code analysis and dynamic app testing need to be covered.

\item \textbf{Serving LLM on mobile. }
Accessing local LLMs is essential in scenarios where internet connectivity is unavailable, yet deploying these ``large'' LLMs on resource-constrained mobile devices poses a challenge. Thus, innovative full-stack solutions are necessary to efficiently serve LLMs on mobile devices.

\item \textbf{Defending against security exploits targeting on-device LLMs.} The attack
surface of LLMs deployed on devices is much larger than those deployed over the
cloud, as the physical on-device LLMs are stored in mobile devices that are
easily accessible to attackers. Therefore, better defending approaches are
required to protect on-device LLMs.

\item \textbf{Providing LLM-powered framework APIs.} Expectedly, mobile apps are
interested in accessing LLMs to enable intelligent features. However, it would
be challenging for app developers to directly interact with LLMs, especially if
they lack the necessary AI knowledge. We therefore argue that there is a need to
provide well-designed framework APIs to facilitate intelligent app development.

\item \textbf{Providing LLM-powered runtime app monitoring.} Recent studies have
presented various runtime monitoring techniques for mobile apps where
provenances are collected and analyzed by remote app vendors to facilitate
runtime profiling, performance optimization, and even mitigating security
exploitations. We anticipate LLMs can offer highly intelligent runtime
monitoring techniques to reason about the provenances and provide insights over
the runtime behavior of mobile apps. Moreover, while recent studies have shown
the potential privacy risks when uploading app logs to remote servers, we note
that LLMs on mobile can be used to analyze these sensitive logs locally without
leaking sensitive information to remote servers.

\end{itemize}

We elaborate on these directions in the following sections.

%Mobile Foundation Model as Firmware The Way Towards a Unified Mobile AI Landscape

\section{Preparing Dataset for Fine-tuning LLMs}

In the domain of software engineering (SE), the preparation of datasets is crucial for the effective training and fine-tuning of LLMs~\cite{Sun_Li_Liu_Du_Li}. Accurate, high-quality, and diverse datasets not only enhance the model's generalization capabilities but also optimize its performance, ensuring reliability in validation and testing. When preparing datasets for fine-tuning LLMs, especially within SE, User Experience (UX), and other multi-modal data processing scenarios, researchers must focus on the collection, classification, preprocessing, and representation of data to ensure its richness and diversity.

\subsection{SE Scenarios}

For SE scenarios, dataset preparation needs to center around specific SE tasks such as code comprehension, bug fixing, code generation, and more. Data sources can be divided into four main categories~\cite{Hou2023LargeLM}: open-source datasets, collected datasets, constructed datasets, and industrial datasets. \textit{Open-source Datasets}~\cite{Chen_Scheurer_Korbak_Campos_Chan_Bowman_Cho_Perez_2023,Khakhar_Mell_Bastani_2023,Wang_Liu_Xie_Li_2023,Zeng_Tan_Zhang_Li_Zhang_Zhang}: Publicly accessible datasets distributed via open-source platforms or repositories. For example, the HumanEval dataset~\cite{Chen_Tworek_Jun_Yuan_Pinto_Kaplan_Edwards_Burda_Joseph_Brockman_et}, containing 164 manually created Python problems with their unit tests. \textit{Collected Datasets}~\cite{Huang_Xia_Xing_Lo_Wang_2018,Mastropaolo_Pascarella_Bavota,Sghaier_Sahraoui_2022,Tian_Lu_Li_Tang_Cheung_Klein_Bissyande_2023}: Datasets compiled by researchers from various sources such as websites, forums, blogs, and social media. Data is often extracted from Stack Overflow threads or GitHub issue comments to tailor datasets for specific research queries. \textit{Constructed Datasets}~\cite{Ezzini_Abualhaija_Arora_Sabetzadeh,Kang_Yoon_Yoo_2022,Koide_Fukushi_Nakano_Chiba_2023,Zhang_Liu_Gong_Zhang_Huang_Jiang_2023}: Datasets specifically designed by researchers by altering or enriching collected data to closely match particular research goals. This includes manually annotating code snippet datasets to study automated program repair technologies, among others. \textit{Industrial Datasets}~\cite{Wang_Shi_Li_Wang_Yang_2020,Alhamed_Storer_2022,Moharil_Sharma_2022}: Comprise proprietary business information, user behavior logs, and other sensitive data from commercial or industrial firms. These datasets are crucial for research targeting real-world business situations but usually require navigating legal barriers to protect commercial interests.

The current research landscape reveals a significant reliance on open-source and collected datasets due to their accessibility and reliability. However, there's a notable gap in the use of constructed datasets (mainly on how are the dataset pre-processed for LLMs) and industrial datasets, indicating a potential disconnect between academic research datasets and those encountered in real-world industrial contexts. Future research directions should aim to bridge this gap by exploring the use of industrial datasets, ensuring that LLMs are applicable and robust across both academic and industrial scenarios.

\subsection{UX Scenarios}   
% \yonghui{what about expand this subsection to human-centric? Involves personalization and adaption to individual users' preferences.   Accessibility, Inclusivity, Education,...}

In UX scenarios, towards improving user experience of using mobile devices, one imperative task is to identify the list of scenarios that can be powered by LLMs.
To achieve this, it requires to
prepare suitable datasets (e.g., diverse user-system interaction data) to train and fine-tune LLMs. Key data sources include the followings. User \textit{Interaction Logs}: Records of user actions within software, websites, or apps, which provide insights into behavior patterns, task workflows, and interface pain points. The goal is to extract significant behavior features and identify any inefficiencies. \textit{User Feedback and Reviews}: Comments from social media, forums, and review systems, which offer valuable perspectives on user satisfaction and expectations. NLP techniques are used to derive sentiments, pinpoint common problems, and gather improvement suggestions. \textit{User Surveys and Interviews}: These direct sources reveal user needs and preferences. The challenge lies in converting responses into a structured format for LLM learning, necessitating careful coding and categorization. \textit{User Testing and Experiments}: Conducted in controlled settings, this data shows how design choices affect user behavior and satisfaction. It's crucial for understanding the impact of different interface designs and functionalities.

In streamlining the discussion on personalization and adaptation in UX scenarios for LLMs, we focus on the essence of crafting user-centric software solutions.  The process hinges on analyzing User Interaction Logs, Feedback and Reviews, and insights from Surveys and Interviews to tailor experiences that resonate with individual preferences.  By dynamically adjusting content and interactions based on a deep understanding of user behaviors and patterns, software can offer a more personalized journey, enhancing user engagement and satisfaction.

The challenge lies in balancing personalized experiences with privacy and security, ensuring data is handled with care.  Moreover, adaptation goes beyond customization to evolve with user feedback and subtle cues, like device type or location, to anticipate and meet unexpressed needs, thereby fostering a deeper connection with the user.

Despite the hurdles of privacy concerns, bias mitigation, and technological limitations, the goal is to develop LLM-powered applications that are not just functional but intuitive and engaging.  This condensed narrative underscores the importance of personalization and adaptation in moving towards more human-centric, responsive, and ultimately more effective software solutions.

\subsection{Multi-Modal Data Processing Scenarios}

Further to the above directions, it is also essential to consider handling multi-modal data available in mobile devices. To date, large models have demonstrated emerging capabilities in handling tasks over multi-modal data, such as text, image, audio, etc. Importantly, deploying LLMs in mobile devices offers multi-modal data exposures, as modern mobile platforms can face various types of domain-specific data from users (e.g., text, photos, audio), sensors (e.g. accelerometer, gyroscope, GPS), wearable devices (e.g., heart rate, sleep quality), and network. Recent studies have shown that LLMs can be fine-tuned to comprehend 
% \lili{XXXX} 
textualized signal
collected from sensors~\cite{xu2024penetrative}. Nevertheless, the integration of LLMs with multi-modal data processing in mobile devices remains largely unexplored. We envision key challenges coming from numerous data sources, data formats, and data types, which require innovative approaches to process and analyze.

More importantly, we envision the possibly of instructing LLMs to process various logs and traces generated by mobile apps and even the mobile operating system (OS) itself. We aim to leverage LLMs to analyze those logs and traces to facilitate runtime profiling, debugging, and performance optimization (see further technical details and discussions in Sec.~\ref{sec:runtime}). Moreover, we anticipate the technical solutions for runtime detection of security exploitations, penetrations, and other anomalies of apps and OS using LLMs. Supporting this vision, we advocate the community to provide datasets that include logs, traces, and other dimensions of provenances to enable proper
fine-tuning and calibration of LLMs.

%which facilitates improving user experience and the system
%performance. Therefore, LLMs are expected to be fine-tuned to process
%multi-modal data in mobile devices, and also support processing logs and traces.

\section{Applying LLMs for Mobile App Development and Analysis}
\label{sec:llm4se}

This section proposes a holistic framework that utilizes the advanced capabilities of LLMs to address critical aspects of mobile app development and analysis. By seamlessly integrating LLMs into processes such as app development, code analysis, app testing, privacy evaluations, and app market analysis, we aim to ensure a secure, user-centric, and optimized digital ecosystem. We now detail a vision where LLMs empower stakeholders across the mobile app landscape, enhancing every facet from code integrity to market dynamics.

\noindent\textbf{Requirements Engineering for Mobile Apps.}
In the specific context of mobile app development, LLMs can revolutionize requirements engineering by automating the translation of user needs into clear, actionable requirements tailored for mobile platforms. They enhance communication among stakeholders, which are crucial for capturing the unique demands of mobile users. LLMs are instrumental in crafting precise documentation and use cases that reflect the mobile user experience, taking into account the constraints and capabilities of mobile devices. They aid in prioritizing requirements with a focus on mobile-specific features and performance expectations. Furthermore, LLMs facilitate the validation process by ensuring requirements are complete and consistent, significantly minimizing the risk of expensive modifications during the critical stages of mobile app development and aligning the project closely with mobile user expectations and project objectives.

% \lili{missing app development tasks}
\noindent\textbf{App Development.}
LLMs could revolutionize the way developers conceive, design, and implement mobile apps. By providing real-time coding assistance, generating code snippets based on developer prompts, and offering optimization suggestions, LLMs can significantly reduce development time and elevate code quality~\cite{bareiss2022code,liu2023your,jiang2023selfevolve,gilbert2023semantic,li2022cctest,pudari2023copilot}. In addition to high-level assistance, LLMs can delve into the intricacies of algorithm optimization, suggesting efficient data structures and algorithms tailored to the app's specific needs and constraints. By analyzing patterns in the developer's coding style, LLMs can offer personalized code refactoring suggestions, ensuring that the codebase remains clean, maintainable, and consistent with the project's architectural principles.
Moreover, through code interpretation, LLMs can elucidate complex code segments, offering clarifications and detailed explanations that enhance developers' understanding of their own and others' code. This leads to improved debugging and maintenance efficiency. The integration of code refactoring capabilities could allow LLMs to suggest structural improvements that increase the readability and performance of the codebase, promoting best practices and design patterns. Additionally, code visualization tools powered by LLMs can transform abstract code structures into intuitive graphical representations, making it easier for developers to grasp the architecture, flow, and dependencies of their applications. These visual aids are instrumental in identifying potential bottlenecks, optimizing workflows, and facilitating collaborative reviews. 

\noindent\textbf{App Code Analysis.}
The core functionality of an app hinges on its complex code, requiring detailed analysis to ensure performance and security. LLMs provide powerful, comprehensive analysis beyond traditional methods~\cite{guo2024exploring,li2022auger,liu2024reliability,lu2023llama,sghaier2023multi,tufano2022using,zhang2022coditt5}.
For example, LLMs can improve static code analysis to thoroughly inspect code without running it, identifying complexities, compliance with coding standards, and risky API uses~\cite{hao2023ev,sun2023dexbert}. This proactive analysis is pivotal in identifying security vulnerabilities, code smells, and performance bottlenecks, effectively preempting issues before they escalate into more significant problems.
LLMs can also enhance code clone detection by analyzing code's syntax and semantics to identify duplicates across apps~\cite{khajezade2024investigating,chochlov2022using,dou2023towards,jiang2023nova}. This could help prevent app cloning, protect originality, and avoid licensing issues, preserving the app ecosystem's integrity.
Furthermore, LLMs can help evaluate third-party libraries in app development, assessing their security, updates, and compatibility. This ensures the integration of only secure and well-maintained libraries, enhancing app security and functionality.
LLMs also play a crucial role in automated program repair~\cite{cao2023study,charalambous2023new,deligiannis2023fixing,fan2022automated,gao2023constructing,huang2023chain,ibrahimzada2023automated,islam2024code}, suggesting fixes for bugs and vulnerabilities, thereby speeding up debugging and enhancing code robustness. Nevertheless, despite extensive research in software engineering, there remains significant room for improvement in the field of mobile app code analysis.

\noindent\textbf{App Testing and Optimization.}
Achieving a seamless and faultless app experience necessitates a relentless pursuit of perfection through rigorous testing and constant optimization. LLMs are revolutionizing this process by automating various facets of testing and optimization~\cite{wang2024software,siddiq2023exploring,li2023finding,yuan2023no,li2023nuances,wu2023large,yang2024large}. In GUI testing~\cite{yoon2023autonomous,liu2022blank}, for instance, LLMs can automate the generation of test cases, predict potential user interactions, and validate UI elements for accessibility and usability standards. This automation extends to bug replay and fixing~\cite{huang2024crashtranslator,kang2022large,kang2023evaluating}, where LLMs can intelligently suggest corrections and optimizations for identified issues, reducing the manual effort required from developers. Moreover, LLMs can optimize app performance by analyzing usage patterns and resource consumption, suggesting efficient algorithms, and predicting user behavior to preload resources or functionalities. This level of automation and insight not only accelerates the development cycle but also ensures that the final product stands up to the highest standards of quality, performance, and user satisfaction.

\noindent\textbf{Privacy-related Analysis.}
As digital privacy~\cite{tang2023user,hamid2023study,jain2023atlas,pan2023seeprivacy} becomes increasingly paramount, LLMs offer a novel approach to navigating the complexities of privacy policies and compliance. By demystifying privacy policies through data mining and ensuring that apps adhere to regulatory standards, LLMs could play a crucial role in fostering a transparent and trust-based relationship between apps and their users.

\noindent\textbf{App Market Ecosystem Analysis.}
In the ever-changing landscape of the app market~\cite{Zhu_2024}, staying abreast of trends and competitive dynamics is key to success. LLMs can offer unparalleled insights into market movements, user preferences, and competitive strategies, empowering developers and marketers to make informed decisions that drive growth and innovation. For example, the voice of the user, encapsulated in reviews, holds invaluable insights into the app experience. Harnessing LLMs to mine this data, developers and researchers can extract pivotal information, classify sentiments, and detect spam with higher accuracy~\cite{ghadhab2021augmenting,kou2023automated,yang2022aspect}. This not only amplifies the value derived from user feedback but also equips developers with the tools to prioritize enhancements and foster an engaging user experience.

% \section{Minimizing LLM for On-device Deployment}
\section{Serving LLM on Mobile}
LLMs have revolutionized NLP tasks with remarkable success on general tasks.
With growing concerns over data privacy and the stringent response latency requirement, running the LLM on mobile devices locally has attracted attention from both academia and industry.
However, their formidable size and computational demands present significant challenges for practical deployment on resource-constrained mobile devices.
This section exclusively focuses on techniques that can be applied to pre-existing LLMs with minimal training efforts, up to the level of fine-tuning, rather than delving into the complexities of designing hardware and models specifically tailored for mobile devices.
Accomplishing full-stack on-device inference optimization necessitates a comprehensive approach that takes into account various aspects of the model, hardware, software, and deployment stack.
Among these optimizations, model-level optimization (model compression) is often considered the most crucial for deploying LLMs on mobile devices.

Model Compression techniques have been intensively investigated to reduce the LLM size and computational complexity without significantly impacting its performance. 
We categorized 4 model compression techniques as detailed in the following, including \textit{Pruning}, \textit{Knowledge Distillation}, \textit{Quantization}, and \textit{Low-rank Factorization}. 
\textbf{Pruning} is one extensively studied technique~\cite{lecun1989optimal, han2015learning, li2016pruning} for removing non-essential components in the model.
Based on removing entire structural units or individual weights, \textit{Pruning} can be divided into Structured Pruning~\cite{anwar2017structured, fang2023depgraph} or Unstructured Pruning~\cite{zhang2018systematic, gordon2020compressing}, respectively, both of which target weight reduction without modifying sparsity during inference.
Contextual pruning~\cite{liu2023deja, valicenti2023mini} differs from the above by its dynamic nature, adjusting the model in real-time based on the context of each inference task. 
\textbf{Knowledge Distillation} (KD)~\cite{hinton2015distilling, kim2016sequence, tung2019similarity} enables the transferring of knowledge from a complex model (LLMs), referred to as the teacher model, to a simpler counterpart known as the student model for deployment. 
Most previous approaches were adopting white-box distillation~\cite{jiao2019tinybert, sanh2019distilbert, sun2019patient}, which requires accessing the entire parameters of the LLM. Due to the arising of API-based LLM services (e.g., ChatGPT), black-box distilled models attract lots of attention, such as Alpaca~\cite{taori2023stanford}, Vicuna~\cite{chiang2023vicuna}, WizardLM~\cite{xu2023wizardlm}, and so on~\cite{peng2023instruction, zhu2023minigpt}.
\textbf{Quantization} has emerged as a widely embraced technique to enable efficient representation of model weights and activations~\cite{liu2021post, gholami2022survey, guo2020accelerating} by transforming traditional representation (floating-point numbers) to integers or other discrete forms.
According to the timing of the quantization process, it can be categorized into post-training quantization (PTQ)~\cite{liu2021post, nagel2020up, fang2020post} and quantization-aware training (QAT)~\cite{tailor2020degree, kim2022understanding, ding2023parameter}. 
\textbf{Low-Rank Factorization}~\cite{cheng2017survey, povey2018semi, idelbayev2020low} is a model compression technique that aims to approximate a given weight matrix by decomposing it into two or more smaller matrices with significantly lower dimensions.

Beyond model compression, the use of the LLM on mobile devices can be further improved through other inference optimizations, which involve \textit{Parallel Computation}, \textit{Memory Management}, \textit{Request Scheduling}, \textit{Kernel Optimization}, and \textit{Software Frameworks}. 
\textbf{Parallel Computation}~\cite{shoeybi2019megatron, pope2023efficiently, borzunov2022petals} leverages modern hardware's parallel processing capabilities to distribute computation across multiple cores or devices, substantially speeding up inference. It can be categorized into model parallelism~\cite{shoeybi2019megatron, pope2023efficiently, narayanan2021memory} and decentralized inference~\cite{borzunov2022petals, borzunov2024distributed, jiang2023hexgen}, depending on the target object being distributed.
\textbf{Memory Management}~\cite{kwon2023efficient, miao2023specinfer, song2023powerinfer} refers to allocating, organizing, and efficiently utilizing the available memory resources on a mobile device.
The Key-Value (KV) cache is a prime optimization target for autoregressive decoder-based models due to the memory-intensive nature of transformer architectures and the need for long-sequence inference~\cite{kwon2023efficient, sheng2023s, zheng2023efficiently}. 
\textbf{Request Scheduling}~\cite{han2022microsecond, ali2020batch, ng2023paella}, similar to general ML serving techniques, aims to schedule incoming inference requests, optimize resource utilization, guarantee response time within latency service level objective (SLO), and effectively handle varying request loads.
Common aspects involve dynamic batching\cite{ali2020batch}, preemption\cite{han2022microsecond}, priority~\cite{ng2023paella}, swapping~\cite{bai2020pipeswitch}, model selection~\cite{gunasekaran2022cocktail}, cost efficiency~\cite{zhang2019mark}, load balancing and resource allocation~\cite{weng2022mlaas}.
\textbf{Kernel Optimization}~\cite{NVIDIAEffectiveTransformer, shi2023welder, zhai2023bytetransformer} focuses on optimizing the individual operations or layers within the model by leveraging hardware-specific features and software techniques to accelerate critical computation kernels. Common aspects involve kernel fusion~\cite{wu2021tentrans}, tailored attention~\cite{lefaudeux2022xformers}, sampling optimization~\cite{fan2018hierarchical}, variable sequence length~\cite{zhai2023bytetransformer}, and automatic compilation~\cite{katel2022mlir}.

\textbf{Software Frameworks}~\cite{MLC-LLM, NVIDIAEffectiveTransformer, FlexGen} play a crucial role in inference optimization by encapsulating complex patterns, practices, and functionalities into reusable high-level APIs or automatic processes, providing abstractions to leverage various techniques for enhanced performance, scalability, and resource utilization.  
Integrating a \textbf{Deep Learning (DL) Compiler} into the framework further streamlines the optimization process with a unified environment for development, optimization, and deployment~\cite{li2020deep}.
The DL compiler takes trained models as input and translates them into optimized code or instructions, often represented as multi-level intermediate representations (IRs), specifically tailored for target hardware platforms, such as CPUs, GPUs, TPUs, or other accelerators. It further applies various analyses and optimization techniques to achieve frontend and backend optimization, resulting in improved performance and efficiency during inference~\cite{chen2018tvm, david2021tensorflow, lattner2020mlir}.
Recent research also offers emerging compiler-aided security hardening techniques to protect the compiled model code~\cite{chen2023obsan}.
Overall, the synergy between software frameworks and DL compilers simplifies the development process, enabling automatic optimization, hardware adaptation, portability, interoperability, and enhanced performance.
By incorporating various advanced techniques, software frameworks offer a pragmatic strategy for boosting inference performance, scalability, and resource utilization, facilitating the development, optimization, and deployment of LLM serving on mobile.

The optimization techniques described are not standalone solutions but are often used together to achieve the best on-device inference performance. 
Additionally, refining LLM inference involves balancing model accuracy with optimizing model size, computational demands, and overall performance, presenting a complex challenge that requires careful consideration.
Beyond striving for efficiency, ensuring the security and protection of the model's intellectual property (IP) adds another layer of intricacy to the optimization efforts. These aspects, along with their implications for the optimization process, will be further discussed in the following on-device LLM security and LLM-Powered frameworks sections.
% \mingyi{I will add a part to introduce the AI compiler here}
% \yonghui{Hi Mingyi, please consider inputting your AI compiler detail within software frameworks related optimization.}

\section{On-device LLM Security}

% \lili{mingyi}

DL techniques such as LLM are deeply engaged in human life. We can use them to revise the article, provide daily recommendations, write codes, and generate image or text content. The data collection required for cloud LLM presents obvious privacy issues. Users’ personal, highly sensitive data have to be shared with computing servers~\cite{shokri2015privacy}. This may cause sensitive information leakage or violate data protection laws~\cite{zhou2023modelobfuscator, zhou2024investigating}. 
Therefore, deploying DL models directly on devices has gained popularity in recent years. However, recent studies show that on-device DL deployment also has serious security issues, especially for LLM. As such DL models are directly hosted on mobile systems, attackers can easily unpack the mobile Apps to obtain the deployed models~\cite{zhou2024investigating}. Because the model weights are trained by a large amount of training data and have extremely high values~\cite{floridi2020gpt,achiam2023gpt}, deploying LLM on devices is a high-risk decision for developers. In addition, the internal information of on-device LLM can be considered a white box for attackers. Even if developers adopt some protections to resist parsing the model information, attackers still can locate the model information and reverse engineer the model details, \ie weights and structure~\cite{zhou2024investigating}. 
%\todored{do we want to add some side channel attacks and hardware fault injection attacks (controlling model outputs and leaking weights)? they can be mitigated by obfuscation as well}
Moreover, recent side-channel attacks and hardware fault injection attacks
(e.g., Rowhammer attacks~\cite{mutlu2019rowhammer}) can also be used to exploit deep learning models, even in the advanced transformer
architectures~\cite{zheng2023trojvit,rakin2020tbt}. For instance, it is shown that these system-level or hardware-level attacks can manipulate the model outputs~\cite{hong2019terminal} by performing Rowhammer attacks to flip certain critical bits in the model weights. Moreover, with the help of queries to the model, attackers can also leak the model weights~\cite{rakin2022deepsteal}.

To protect the deployed DL models, especially for LLM, we now have two main methods to defend the on-device models: Trusted Execution Environments (TEE) and program protection. For the Trusted Execution Environments (TEE)~\cite{lee2020keystone,wagh2020falcon,mo2020darknetz,chen2020training,mo2021ppfl,zuo2021sealing}, it provides secured execution environment for on-device models. These methods design customized software or hardware architecture for protect the ownership of the deployed model, disable the access of unauthorized parties, and generate an encrypted model inference pipeline. These methods are effective in protecting the deployed model. However, they are hard to apply to various mobile platforms such as Android because they usually need specially designed software or hardware architectures. In addition, attackers are capable of using side-channel attacks to infer the model architectures~\cite{wei2018know,xiang2020open,batina2019csi,rakin2022deepsteal,yu2020deepem,wei2020leaky,zhang2021stealing}.

\begin{figure}[!h]
    \centering
    \includegraphics[width=0.85\linewidth]{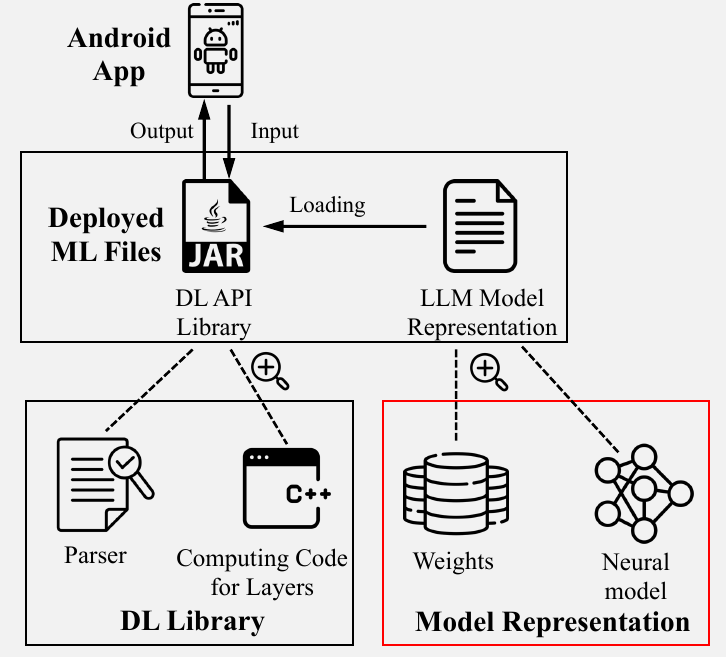}
	\caption{The information leakage problem of on-device LLM on Android. The sensitive model representation is directly hosted on mobile devices.}
    \label{fig:llm_sec_overview}
\end{figure}

To protect the LLM on various mobile systems and devices, model protection can be considered a special program protection problem. The general protection
method for software such as obfuscation and optimization can also be applied to LLM on mobile. As shown in Figure~\ref{fig:llm_sec_overview}, the security issue of on-device LLM is mainly caused by the exposure of the model representation (the red block of Figure~\ref{fig:llm_sec_overview}). Attackers can reverse engineer the model representation that is packed in the deployed AI programs, \eg model files and API libraries, to steal the intellectual property~\cite{zhang2018protecting,xue2021intellectual} or generate effective white-box attacks~\cite{nasr2019comprehensive,zhou2020dast,zhang2022investigating,huang2022smart,zhou2024investigating}. Therefore, minimizing the exposure of model representation can effectively protect the on-device LLM. To this end, Zhou~\etal~\cite{zhou2023modelobfuscator} adopt the idea of code obfuscation~\cite{schrittwieser2016protecting,collberg1997taxonomy,collberg1998manufacturing,wang2001security,wroblewski2002general}, which is a well-developed approach for hiding sensitive information in software, and propose to obfuscate the information of on-device ML models. Like the obfuscated code, the obfuscated on-device model contains hard-to-read information but still can be correctly run on the mobile devices. It can significantly increase the difficulty of reverse engineering the deployed LLM. In addition, a program refactorization scheme has been proposed to hide the explicit model representation on devices~\cite{zhou2024modelless}. Unlike the other tools that only support limited number of model architectures and formats like \textit{m2cgen}\footnote{\href{https://github.com/BayesWitnesses/m2cgen}https://github.com/BayesWitnesses/m2cgen} and \textit{llama.cpp}\footnote{\href{https://github.com/ggerganov/llama.cpp}https://github.com/ggerganov/llama.cpp}, it automatically trace the function call of model inference, extract the related codes, and refactor the code into an executable program. This scheme can applied to commonly used DL models such as LLM. The generated program does not have explicit model representation, \ie model weights and architecture. Attackers need to use human efforts to understand the compiled binary file to reverse engineer the deployed models. Accordingly, given the model becomes much obscure and hard to analyze, side channel attacks and hardware fault injection attacks are also hard to be applied to the protected models to achieve high attack accuracies (e.g., the target critical model weights are hard to localise and manipulated)~\cite{rane2015raccoon}.

Overall, although defense strategies based on program protection can be applied to almost all mobile platforms, it is worthy noting that these strategies cannot disable the reverse engineering of on-device LLM. Their goal is to significantly increase the cost of attackers, \ie using lots of human efforts to understand the binary program. The TEE-like defense methods are more suitable to be applied to high-value systems. In contrast, the program defense strategy can be applied to various Apps on various mobile OS.

\section{Providing LLM-powered Framework APIs}

% \lili{Build Lightweight Longchain-alike frameworks}

% \lili{chen xiao}
The exploration of LLM-powered framework APIs for mobile app development is a vibrant and expanding field, focusing on streamlining the integration of advanced language models into mobile applications. This area of research is dedicated to the development, optimization, and deployment of APIs that enable mobile apps to leverage the capabilities of LLMs for a wide range of tasks, including natural language processing, conversational interfaces, and content generation.

Recent advancements have concentrated on creating accessible, efficient, and scalable solutions \cite{LangChain,FlowiseAI,GradientJ}. Frameworks are being developed to simplify the integration of LLMs into various applications, offering APIs that abstract away the complexities of direct interactions with LLMs. This makes it easier for developers to implement advanced language capabilities in their applications. Additionally, these frameworks are evolving to support more context-aware interactions, allowing LLMs to provide more relevant and personalized responses based on the user's context and previous interactions with the app \cite{LangChain}.

Looking forward, the functionality and utility of LLM-powered framework APIs for mobile app development could be significantly enhanced through focused research in several key areas. The development of standardized API protocols promises to facilitate a more uniform development experience across different mobile operating systems and device types. Standardizing APIs could ensure that LLM-powered features are consistently available across the mobile ecosystem, catering to the diverse needs of developers and users alike.

Security is another critical area requiring attention. As the integration of LLMs into mobile apps increases, addressing the security implications of these APIs becomes imperative. Future research will need to explore ways to ensure secure data transmission between mobile devices and cloud servers, as well as secure on-device processing to minimize data exposure. This will be crucial in maintaining user trust and protecting sensitive information.

Energy efficiency is also a critical concern, given the limited battery life of mobile devices. Future directions should include research into mechanisms for minimizing the energy consumption of LLM-powered APIs. This could involve developing smarter caching strategies or optimizing the computational workload distribution between the device and the cloud, ensuring that mobile applications can deliver advanced functionalities without excessively draining battery life.

Additionally, the potential for LLM-powered APIs to support more interactive and multimodal inputs, such as combining text, voice, and visual inputs, opens up interesting new possibilities. This evolution could enable more natural and engaging user interactions with mobile applications, creating new possibilities for app design and functionality. Such advancements would not only enhance the user experience but also pave the way for innovative applications that fully exploit the capabilities of LLMs.

\section{Providing LLM-powered Runtime Monitoring}
\label{sec:runtime}

%\noindent\textbf{App Runtime Analysis.}~

Further to the above directions, LLMs can be deployed to monitor the runtime
behavior of mobile apps for various software engineering and security purposes.
This is particularly important given the increasing complexity of mobile apps
and the potential security threats they face; for instance, mobile apps can be
attacked to leak sensitive user information, disrupt services, or even
compromise the mobile device. Nevertheless, offline analysis and testing of
mobile apps' behavior may be likely insufficient to detect and prevent all those
runtime attacks. From this perspective, we envision that LLMs can be deployed in
mobile devices to monitor the runtime behavior of mobile apps, the mobile
frameworks, and even the mobile operating system (OS) itself for various
software engineering and security purposes.

\noindent \textbf{Offering Intelligent Runtime Analysis.}~LLMs have demonstrated
state-of-the-art performance in a wide range of natural language and code
processing tasks. In particular, it is shown that LLMs can reason real-world software
artifacts and other complex scenarios, given that they have been trained on
large-scale corpora which often subsume common sense knowledge and programming
expertise. With the high reasoning capability, we envision that LLMs can be
deployed to monitor the runtime behavior of mobile apps to facilitate various
software engineering tasks, such as profiling, debugging, and performance
optimization. Furthermore, given that possible attacks can be launched against
mobile apps and even the mobile frameworks, we see that LLMs can be deployed to
monitor and reason the runtime behavior and recognize potential security
threats. To enhance the intelligence of LLMs in analyzing those collected
information, we envision that LLMs can be fine-tuned with relevant trace
datasets to better reason the runtime behavior of mobile apps; we also expect
LLMs to incorporate domain-specific knowledge of common security threats
encountered by mobile apps. Prompt engineering techniques like chain-of-trust
can also be adopted in this context. Overall, we see the high potential of LLMs
to behave as a ``smart'' runtime analysis system for mobile apps, which can
provide insights into the runtime behavior of mobile apps and the mobile system
and outperforms traditional runtime analysis tools.

\noindent \textbf{Offering Privacy-Preserving Runtime Analysis.}~To facilitate
app vendors to continuously analyze the released mobile apps, the common
practice is that mobile apps generate runtime logs (e.g., crash reports and
traces) and upload them to remote servers for further analysis. This practice is
widely used in real-world scenarios, yet it raises privacy concerns as the logs
may contain sensitive user information. In fact, recent studies have shown the
potential privacy risks of logs and traces generated by mobile apps, which can
leak sensitive user information like doctor
appointments~\cite{hao2021differential}. While some privacy-preserving
techniques have been proposed to sanitize logs and traces before uploading them
to remote servers~\cite{hao2021differential,wang2024ppsca}, they essentially
undermine the utility of logs and traces for further analysis. Moreover, the
mainstream approaches rely on differential privacy techniques, which only offer
limited privacy guarantees and may not be sufficient to protect group users'
privacy and confidentiality. While some advanced techniques like secure
multi-party computation (MPC) and anonymized transmissions may be be used to
enable remote vendor analysis without leaking sensitive information, they are
often computationally expensive and impose a high requirement on the computing
resources on mobile devices. From this perspective, we believe that with LLMs
deployed in mobile, app logs can be analyzed for most cases without leaking
sensitive information to the remote vendor servers. This offers a principled way
to protect user privacy; before releasing the mobile app, the app vendor can
configure the LLMs in the mobile such that the LLMs can better analyze the logs
locally to decide performance issues or security threats. LLMs can analyze the
raw logs to decide performance issues or security threats, and query the remote
vendor servers only when necessary to obtain further insights. This way, the
sensitive information in the raw logs will not be leaked to the remote vendor
servers, and the user privacy will be protected. 

\noindent \textbf{Design Considerations.}~To facilitate such demanding runtime
analysis, we expect to conduct the following tasks. On one hand, this requires
the mobile apps and mobile system components under protection to provide proper
logs and introspection interfaces. LLMs can hook the provided interfaces to
capture the runtime behavior of mobile apps, and even the mobile frameworks and
the mobile OS. 
Interestingly, instead of forming a ``passive'' runtime analysis system where
LLMs wait for logs and traces to be generated, we envision that LLMs can be
trained to actively interact with mobile apps and the system software to perform
investigation. For instance, once the LLM detects a potential security threat,
it can interact with the mobile app to further confirm the threat and then
decide to take corresponding actions like alerting the user or even terminating
the app. This shall offer a more proactive and efficient runtime analysis system
for mobile apps.
On the other hand, we anticipate the demand of fine-tuning LLMs for such
security tasks. Our tentative exploration shows that mainstream LLMs available
on the market are not sufficiently trained with software trace data, which is
crucial for runtime analysis. Therefore, we advocate the community to provide
relevant datasets to support LLM fine-tuning and customization for runtime
monitoring and analysis tasks.

Recent research has illustrated the high feasibility of using LLMs in relevant
fields~\cite{li2023exploring,jiang2024lilac}; this indicates the high potential
of using LLMs for mobile runtime analysis for software engineering and security
purposes. However, there still exist several challenges to be addressed in the
context of mobile. For instance, we see the demand of augmenting the LLMs'
response time to avoid noticeable delays in mobile devices. More importantly, we
envision the need for ensuring the LLMs' robustness against even privileged
adversaries with access to the device or the LLM model itself. One may also need
to consider the potential ``memorization'' issues of LLMs, which may lead to
cross-app privacy leakage when malicious apps are installed on the same device
and exploit the LLMs' memorization capabilities. We believe that addressing
these challenges will pave the way for deploying LLMs in mobile devices for
runtime analysis tasks. 

\section{Conclusion}

In this position paper, we have motivated the strong necessity to apply LLMs for the mobile ecosystem and subsequently provided an initial roadmap for our fellow researchers to achieve that objective.
In the roadmap, we summarized six directions that we believe are urgently required to be researched, including (1) preparing more datasets, (2) Addressing MSE tasks, (3) Serving LLM on mobile (4) Enhancing the security of on-device LLMs, (5) facilitating intelligent app development through LLM-powered framework APIs, and (6) providing LLM-powered runtime monitoring.
We acknowledge to the community that, these six directions should not be considered as representative to the whole space of applying LLMs for mobile. We would like to invite our fellow researchers to help in identifying more research gaps that need to be filled in order to achieve intelligent user experiences.

\bibliographystyle{ACM-Reference-Format}
\bibliography{acmart}

\end{document}